# Condensedly: comprehending article contents through condensed texts

Chao-Hsuan Ke[1,3], Tsung-Lu Michael Lee[2] and Jung-Hsien Chiang[1,*]

[1]Department of Computer Science and Information Engineering, National Cheng Kung University, Tainan 70101, Taiwan
[2]Department of Information Engineering, Kun Shan University, Tainan 71003, Taiwan
[3]Delta Research Center, Delta Electronics, Inc., Taipei 11491, Taiwan

**ABSTRACT**
**Summary:** Abstracts in biomedical articles can provide a quick overview of the articles but detailed information cannot be obtained without reading full-text contents. Full-text articles certainly generate more information and contents; however, accessing full-text documents is usually time consuming. Condensedly is a web-based application, which provides readers an easy and efficient way to access full-text paragraphs using sentences in abstracts as fishing bait to retrieve the big fish reside in full-text. Condensedly is based on the paragraph ranking algorithm, which evaluates and ranks full-text paragraphs based on their association scores with sentences in abstracts.
**Availability:** http://140.116.247.185/~research/Condensedly

## 1 INTRODUCTION

Comparing the readability of abstracts and full-text articles, abstracts provide definite and concise information on the content of an article, but its brevity inevitably entails the loss of detail (Cohen, *et al.*, 2010). It was obvious that searching in a full-text article is more likely to find relevant articles than searching solely on the abstract (Lin, 2009). Recently, more full-text scientific literatures have become available on the web. They provide users from searching bibliographic records to directly access full-text contents. It has become easier to keep track of research topic or sufficient contents by searching literature online.

Identifying key concepts and detailed descriptions within full-text articles can be done with the assistance of text mining algorithms and programs. So far, several studies have been conducted to explore the article concepts. BioTextQuest clusters significant terms in articles and offers visualization interface to illustrate the concepts. It assists users in identifying the association concepts among topic-related articles (Papanikolaou, *et al.*, 2011). BioIE has implemented a rule-based method to extract specific categories of sentences and generate association between protein function and diseases (Divoli and Attwood, 2005). An alternative approach is the summarization of large query articles using a rule-based method and a semantic relation analysis. The important portions of the article are extracted by relying on the presence of significant terms or concepts and makes the full-text article more streamlined. To sum up, the summary information can benefit researchers to save time in querying in large articles, or other text mining-related tasks such as data curation and text annotation by removing redundant information.

Condensedly is a web application that generates abstract-related condensed sentences ranked from full-text content of an article, which is available at the PubMed Central (PMC). The computational method is based on the paragraph ranking algorithm (Chiang, *et al.*, 2011), a prototype to recommend and rank significant paragraphs which are related to specific topics mentioned in the abstract. It collects these paragraphs and turns them into condensed texts to readers. Condensed texts are expected to provide supplementary information to the abstract content, and sentences within the abstract are considered crucial to condense a research paper. In our previous study, abstract-related paragraphs made up >80% of important paragraphs, and 96% of paragraphs labeled with most important level were considered to be abstract-related. Our results suggest that taking advantage of information in the abstracts could contribute to the retrieval of important paragraphs from full-text articles.

## 2 SYSTEM AND FUNCTIONALITY

In Condensedly, there is a web interface for querying and accessing full-text articles and their condensed texts. Condensed sentences were extracted from full-text paragraph and each abstract sentence has its associated condensed texts. Various bio-entities are highlighted in colors to provide more intuitive access to the contents. S1 (in supplementation) illustrated the search results of our system, Condensedly. The web interface of the system is separated into two panels. The main panel on the right shows the article and author's information. Except for the abstract and the full-text articles, the generated condensed text was also shown in this panel. In addition, users can click on each sentence in the abstract to see its corresponding paragraphs in detail. The panel on the left provides an overview of all bio-entities found in the article, which were sorted by their overall frequency. Condensedly has been tested in four common web browsers, including the Microsoft Internet Explorer 11, Mozilla Firefox 36, Google Chrome 41and Apple Inc. Safari 8.0. The complete function descriptions in Condensedly are described as below.

### 2.1 Search function

To query biomedical articles, a Google-like search function flexibly supports user input keywords, PMID, or a combination of bio-entities with Boolean operations (AND, OR and NOT). Alternatively, we used Lucene (http://lucene.apache.org/) as text storage to process all downloaded PMC XML files and is updated on a regular basis. Furthermore, Lucene can also be used as query processor and ranking engine. Until March 2015, Condensedly provides access up to 1 million openly available PMC full texts, the repository contains > 9503,466 recognized bio-entities.

### 2.2 Construction of condensed text

Paragraph ranking algorithm introduced keywords in abstract to describe the topical information of the article. We assume those keywords may constitute useful entries for linking an abstract sentence to specific paragraphs that describes similar or identical concepts/topics, or may serve as a concise summary for a given document. Through keywords, every sentence within an abstract can be associated with full-text paragraphs, which were ranked by paragraph ranking algorithm. Each sentence in an abstract was treated as a query sentence (QS), then a metric named paragraph relevance–inverse sentence relevance (PR–ISR) was used to evaluate which paragraphs were more significant and relevant to QS. Specific paragraphs are related to topics mentioned in the abstract that contain complementary and detailed information. Another important ranking measurement is the relevant section score (RSS), it evaluates the relationship between QS and a given section in an article. We hypothesize that a section is usually emphasizes a few concepts, and the explanation or description of that those concepts cover multiple paragraphs. To avoid condensed texts favoring a specific section, the condensed texts first choose a candidate section and then recommend the most relevant paragraph from the chosen section by RSS and PR–ISR, respectively.

### 2.3 Named entity recognition

Condensedly recognized and highlighted various bio-entities to help users quickly focus on key terms and provide additional information on them. We used a multitude of state-of-the-art named entity recognition (NER) tools optimized for recognizing mentions from eight different entity classes. In Condensedly, full-text articles were annotated automatically with eight different NER software tools. This function assists users in identifying the important conceptual objects within an article. The eight conceptual objects contain gene names, chemicals, diseases, drugs, SNPs, species names, MeSH terms and abbreviation recognition (see supplementary data for detailed bio-entities recognition methods).

## 3 EVALUATION AND USER STUDY

To gain an understanding of how accurately our approach can identify important paragraphs, a keyword association ratio between abstract sentences and full-text paragraphs, named *IO*, is implemented in our previous study (Chiang*, et al.*, 2011). In the past, we conducted a preliminary experiment to evaluate the correlation between *IO* and full-text paragraphs. A total of 3610 paragraphs from 100 full-text articles were evaluated from level 1 (trivial) to 5 (important) which indicated the relative importance of each paragraph. The paragraphs at level 5 had an average *IO* of 0.635, showing the important paragraphs were related to the abstract. The results of other levels have shown that the *IO* and a paragraph's importance were positively correlated. We then evaluated our condensed text using the ROUGE score (Lin, 2004), and obtained an average ROUGE-1 score of 0.646.

In addition, to understand how users responded to our system, we further performed manual evaluation based on the same 100 articles. We invited five experts to judge whether reviewing only condensed text can usefully comprehend the complete concept of an article. The five experts performed human evaluations, they read 20 articles' condensed texts, and confirmed whether the condensed texts can truly comprehend complete expression of the concept rather than reading the entire article. We asked three questions and used a number 1 (bad) to 5 (good) rating scales to record the satisfaction of each condensed text output, and received a rating of 4.25 on average. The supplementary material describes the manual evaluation in detail.

## 4 CONCLUSION

We describe Condensedly, an online text mining system, enabling condensed texts discovery of the biomedical full-text literatures in PubMed Central using the paragraph ranking algorithm. It provides quick emphasis of accessing full-text research articles and summarizes important biomedical information and bio-entities through web services. We believe this web application not only help users to ease the burden of reading full-text contents but also improve the efficiency of other text mining tasks such as gene annotation and biocuration.

### ACKNOWLEDGEMENTS

We would like to thank all the participants for their contribution to the manual evaluation for helpful discussions throughout the preparation of the manuscript. We also would like to thank Jiun-Huang Ju and Yu-De Chen for valuable comments on web implementation of the system.

*Conflict of Interest*: none declared.